\documentclass[prb,twocolumn,aps,showpacs]{revtex4}
\usepackage{graphicx}%
\usepackage{dcolumn}
\usepackage{amsmath}
\usepackage{amssymb}
\usepackage{epsfig}
\usepackage[latin1]{inputenc}
\usepackage{amsfonts}
\include{biblio}
\usepackage{color}
\usepackage{needspace}
\usepackage{makeidx}
\begin{document}
\title{Speed of sound of a spin balanced Fermi gas with s- and d-wave pairings across the BCS-BEC evolution }
\author{Zlatko  Koinov, Rafael Mendoza }\affiliation{Department of Physics and Astronomy,
University of Texas at San Antonio, San Antonio, TX 78249, USA}
\email{Zlatko.Koinov@utsa.edu}
 \begin{abstract}
The authors of a recent paper (PRA \textbf{87}, 013613 (2013)) argued that in fermionic systems with d-wave pairing the speed of sound is nonanalytic across the BCS-BEC crossover at the point where the chemical potential vanishes, regardless of the specific details of the interaction potential. On the contrary, the numerical results reported here  suggest that the speed of sound across the BCS-BEC evolution  of atomic Fermi gases with s- and d-wave pairings in two-dimensional
 square lattices is a smooth analytic function at the vanishing
chemical potential.
\end{abstract}
 \maketitle
\section{Introduction}
 It is widely accepted that  the s-wave superfluidity
of Fermi gases in  two-dimensional optical lattices  can be described at low
energies and temperatures by a BCS theory with an effective on-site attractive interaction between square-lattice atoms
  that leads to a
 s-wave gap $\Delta_s$.  In the mean-field approximation, the  number- and gap-equations for a fixed filling factor $f$ at zero temperature are
\begin{equation}1=\frac{U}{N}\sum_\textbf{k}\frac{1}{2\sqrt{\xi^2(\textbf{k})+\Delta^2_s}},\label{GU}
\end{equation}
 \begin{equation}f=\frac{1}{N}\sum_\textbf{k}\left(1-
\frac{\xi(\textbf{k})}{\sqrt{\xi^2(\textbf{k})+\Delta^2_s}}\right),\label{NU}
\end{equation}
\begin{figure}
\includegraphics[scale=0.7]{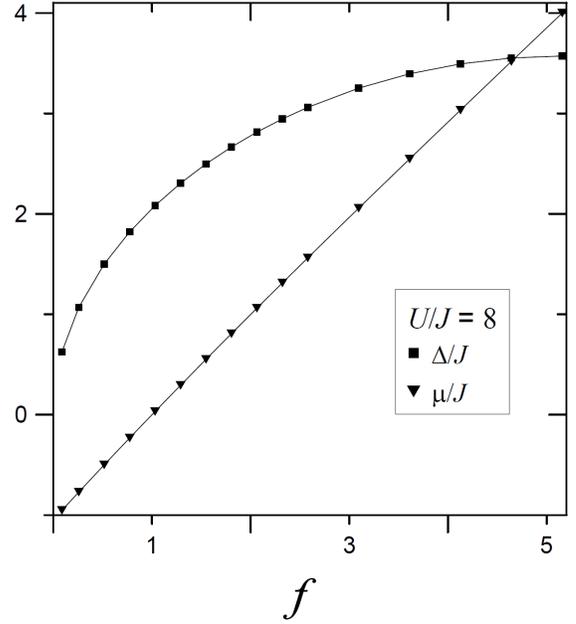}
    \caption{The s-wave pairing gap $\Delta_s$, and the chemical
potential $\mu$ as functions of filling factor $f$ (solid
lines are guides to the eyes). The interaction strength is $U=8J$, and the filling factor has been normalized such that $f=1$ at $\mu=0$.}\label{Fig1}\end{figure}
where $\xi(\textbf{k})=2J(1-\cos k_x)+2t(1-\cos k_y)-\mu$ is the tight-binding  dispersion energy, $J$ is  the tunneling strength of the atoms between
nearest-neighbor sites of the two-dimensional (2D) lattice,  $\mu$ is the chemical potential, and $U$ is the  strength of the attractive interaction (the lattice spacing is
assumed to be $a=1$).

It is evident from many numerical calculations, based on the Gaussian approximation,\cite{G1} the density-response-function approach,\cite{Jap} or the Bethe-Salpeter (BS) formalism,\cite{BS} that in fermion systems with s-wave pairing we have a smooth BCS-BEC crossover. For instance, the collective-mode dispersion within the Gaussian approximation at  zero temperature  is defined by the zeros of the following $2\times 2$ secular determinant:
\begin{equation}
\left|
\begin{array}{cc}
\frac{2}{U}+I^{0,0}_{\gamma,\gamma}+I^{0,0}_{l,l}-2J^{0,0}_{\gamma,l}&I^{0,0}_{l,l}-I^{0,0}_{\gamma,\gamma}\\
I^{0,0}_{l,l}-I^{0,0}_{\gamma,\gamma}&\frac{2}{U}+I^{0,0}_{\gamma,\gamma}+I^{0,0}_{l,l}+2J^{0,0}_{\gamma,l}
\end{array}%
\right|,
\label{GaussU}\end{equation}
where  $I^{0,0}_{\gamma,\gamma},I^{0,0}_{l,l}$ and $J^{0,0}_{\gamma,l}$ are defined in  Section II.

Years ago, it was pointed out that the  Gaussian approximation can be obtained  by summing
diagrams corresponding only to the direct interaction.\cite{Com} Diagrammatically, this means that the Gaussian approximation includes only  contributions
from the ladder diagrams. The  BS formalism, as well as the density-response-function approach, allow us to take into account contributions from both the ladder diagrams (direct interaction) and bubble diagrams (exchange interaction). When both interactions are taken into account, the secular determinant becomes:
\begin{equation}\left|%
\begin{array}{cccc}
 1+UI^{0,0}_{\gamma,\gamma}&UJ^{0,0}_{\gamma,l}&UI^{0,0}_{\gamma,\widetilde{\gamma}}&UJ^{0,0}_{\gamma,m} \\
   UJ^{0,0}_{\gamma,l}&1+UI^{0,0}_{l,l}&UJ^{0,0}_{l,\widetilde{\gamma}}&UI^{0,0}_{l,m} \\
    UI^{0,0}_{\gamma,\widetilde{\gamma}}&UJ^{0,0}_{l,\widetilde{\gamma}}&-1+UI^{0,0}_{\widetilde{\gamma},\widetilde{\gamma}}&UJ^{0,0}_{\widetilde{\gamma},m} \\
     UJ^{0,0}_{\gamma,m}&UI^{0,0}_{l,m}&UJ^{0,0}_{\widetilde{\gamma},m}&1+UI^{0,0}_{m,m} \\
\end{array}%
\right|\label{BSU}
\end{equation}
In Fig. 1, we have shown the solutions of the number- and gap-equations across the BSC-BEC evolution as functions of the filling factor $f$. The interaction strength is  $U=8J$.  The chemical potential vanishes at $f=0.1938$, and we have normalized the filling factor such that $f=1$ at $\mu=0$. Using the s-wave pairing gap $\Delta_s$, and the chemical
potentials $\mu$, presented in Fig. 1, we have calculated the speed of sound along the $(Q_x,0)$ direction as a function of the filling factor $f$. The numerical results are shown in Fig. 2. As shown, the speed of sound is a smooth function across the BCS-BEC evolution. Also shown that the Gaussian approximation overestimates the speed of sound when compared to the BS formalism (or to the  density-response-function approach).
\begin{figure}\includegraphics[scale=0.7]{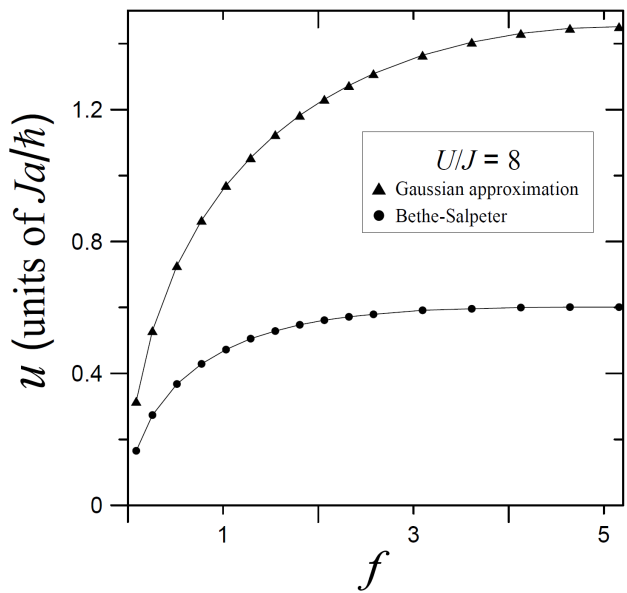}
  \caption{The speed of sound in 2D optical lattice
along the $(Q_x, 0)$ direction as a function of the filling factor $f$ for s-wave pairing calculated within the Gaussian approximation, and by  the BS formalism (solid
lines are guides to the eyes). The interaction strength is $U=8J$, and the filling factor has been normalized such that $f=1$ at $\mu=0$. }\label{Fig2}\end{figure}

 One may well
ask whether the BCS-BEC evolution is smooth crossover in
fermion systems with d-wave pairing. Recently, the BCS-BEC evolution in a 2D fermion system with d-wave pairing has been studied in Ref. [\onlinecite{CHZ}]. The d-wave
pairing is naturally associated with near-neighbor interactions, but  the assumption in Ref. [\onlinecite{CHZ}] is that the short-coherence-length d-wave superfluidity is associated to the separable potential $V(\textbf{k}-\textbf{q})=\lambda\Gamma(\textbf{k})\Gamma^*(\textbf{q})$, where $\lambda$ is the strength of the interaction, and $\Gamma(\textbf{k}\rightarrow 0)\sim k^2$. Under this assumption,  the collective-mode dispersion $\omega(\textbf{Q})$ has been numerically calculated  within   the Gaussian-fluctuation approximation.  This approximation provides a $2\times 2$ secular determinant, and the speed of sound $u$ (the slope of the linear part of the graph $\omega$ vs. $Q$) was obtained by replacing the exact elements of the secular determinant with their Taylor expansions about the points $Q=0$ and $\omega=0$. The result is that the speed of sound depends on a 2D-momentum integral, with this integral being divergent at the point $\mu=0$ for d-wave pairing gap $\Delta_\textbf{k}\sim \Gamma(\textbf{k})$. The divergence of the integral is the reason to argue that the speed of sound  is non-analytic function across the BCS-BEC transition at $\mu=0$, regardless of the specific details of the interaction potential.

In what follows, we report results concerning the speed of sound of a superfluid atomic spin
balanced Fermi gas in a 2D optical lattice across the BCS-BEC evolution assuming   $d_{x^2-y^2}$-wave pairing gap of $\Delta_\textbf{k}=\Delta\left(\cos k_x-\cos k_y\right)/2$, where $\Delta$ is the maximum value of the gap.  d-wave pairing gap has been  observed  in high-$T_c-$cuprates.  The  cuprate superconductors are close to an antiferromagnetic instability  driven by a strong on-site repulsion $U$, and the near-neighbor interactions are the spin independent and attractive $V$,  and the Heisenberg-type antiferromagnetic. In optical lattices, the $d_{x^2-y^2}$-wave gap  is associated with the attractive  interaction between the atoms on the near-neighbor sites of the lattice. Decades ago, it was pointed out that the
phase diagram at half filling of the extended Hubbard model shows an
"island" in U-V space where d-wave pairing exists.\cite{UV} There are no experimental
data suggesting that a quantum phase transition  at vanishing
chemical potential takes place across the BCS-BEC evolution in systems described by the extended Hubbard model.

In this communication, we  evaluate the collective-excitation energies of the extended
Hubbard model in the generalized random phase approximation (GRPA). The Hamiltonian of the extended
Hubbard model contains
two interactions representing the on-site Hubbard  interaction $U$ and the attractive interaction $V$ between the atoms on the near-neighbor sites of the lattice:
\begin{equation}\begin{split}&H=-\sum_{i,j,\sigma}J_{ij}\psi^\dag_{i,\sigma}\psi_{j,\sigma}
-\mu\sum_{i,\sigma}\widehat{n}_{i,\sigma}-U\sum_i
\widehat{n}_{i,\uparrow} \widehat{n}_{i,\downarrow}\\&-V\sum_{<i,j>\sigma\sigma'}\widehat{n}_{i,\sigma}\widehat{n}_{j,\sigma'}.
\label{H2}\end{split}\end{equation}
The Fermi operator $\psi^\dag_{i,\sigma}$ ($\psi_{i,\sigma}$)
creates (destroys) a fermion on the lattice site $i$  with spin
projection $\sigma=\uparrow,\downarrow$ along a specified direction,
and $\widehat{n}_{i,\sigma}=\psi^\dag_{i,\sigma}\psi_{i,\sigma}$ is
the density operator on site $i$.   The symbol $\sum_{<ij>}$ means sum
over nearest-neighbor sites of a square lattice. The first term
in (\ref{H2}) is the usual kinetic energy term in a tight-binding
approximation, where $J_{ij}$ is  the tunneling strength of the atoms between
sites $i$ and $j$.  The total number of sites is $N$, and the number of fermion atoms is $M$ (filling factor $f=M/N$).

 In the presence of the nearest-neighbor interaction, the  gap equation (\ref{GU}) becomes \begin{equation}\Delta_\textbf{k}=\frac{1}{N}\sum_\textbf{q}
[U+V(\textbf{k}-\textbf{q})]\frac{\Delta_\textbf{q}}{2E(\textbf{q})},\label{GGap}\end{equation}
where  $E(\textbf{k})=\sqrt{\xi^2(\textbf{k})+\Delta^2_\textbf{k}}$, and the interaction is $V(\textbf{k})=2\lambda(\cos k_x+\cos k_y)$.  Since  the interaction  can be factorized as $V(\textbf{k}-\textbf{q})=\lambda[\Gamma_1(\textbf{k})\Gamma_1(\textbf{q})+\Gamma_2(\textbf{k})\Gamma_2(\textbf{q})+\Gamma_3(\textbf{k})\Gamma_3(\textbf{q})+
\Gamma_4(\textbf{k})\Gamma_4(\textbf{q})]$, where $\Gamma_1(\textbf{k})=\cos k_x - \cos k_y,  \Gamma_2(\textbf{k})=\sin k_x - \sin k_y, \Gamma_3(\textbf{k})=\cos k_x + \cos k_y, \Gamma_4(\textbf{k})=\sin k_x + \sin k_y$,  the gap is defined as $\Delta_\textbf{k}=\Delta_s+\Delta \Gamma_1(\textbf{k})/2+\Delta_1\Gamma_3(\textbf{k})/2$.    If the Hubbard on-site interaction is weak, the s-wave pairing gap $\Delta_s$ can be neglected. The $d_{x^2-y^2}$-wave pairing in the phase diagram appears  in the limit $\Delta>>\Delta_1$. In this limit, the chemical potential $\mu$ and the maximum value of the gap $\Delta$   are defined by the solutions of the mean-field gap- and number-equations:
\begin{equation}
1=\frac{\lambda}{N}\sum_\textbf{k}
\frac{\left(\cos k_x-\cos k_y\right)^2}{2E(\textbf{k})},
\label{Gp}
\end{equation}
\begin{equation}
f=\frac{1}{N}\sum_\textbf{k}\left(1-
\frac{\xi(\textbf{k})}{E(\textbf{k})}\right).
\label{Num}
\end{equation}
\begin{figure}\includegraphics[scale=0.7]{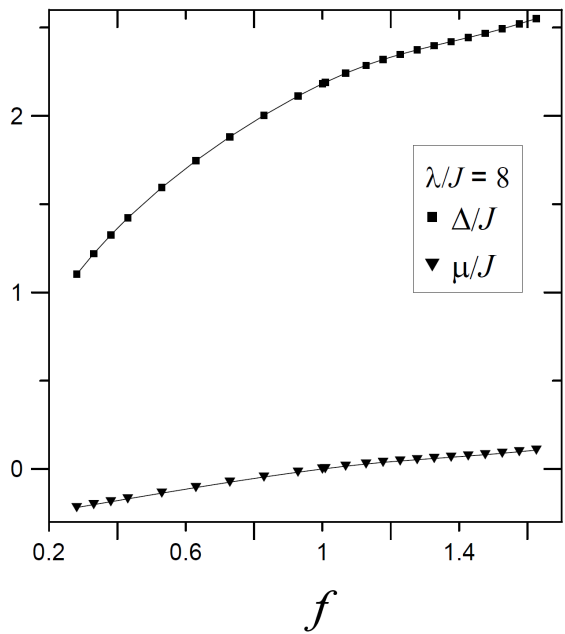}
    \caption{The maximum value of the  gap $\Delta$, and the chemical
potential $\mu$ as functions of filling factor $f$ (solid
lines are guides to the eyes). The interaction strength is $\lambda=8J$, and the filling factor has been normalized such that $f=1$ at $\mu=0$. }\label{Fig3}\end{figure}

The above mean-field
number- and gap-equations were solved at zero temperature. The corresponding mean-field results for the maximum value of the gap $\Delta$ and the chemical potential $\mu$ (in units $J$) are shown in Fig. 3.  In the numerical calculations,  the
strength of the nearest-neighbor interaction is fixed at $\lambda=8J$.  The chemical potential vanishes at a filling factor  $f=0.0502$.

The solutions of Eqs. (\ref{Gp}) and (\ref{Num}) correspond to the minimum of the zero-temperature free energy of the system $F(\mu,\Delta)=\Delta^2/\lambda+1/N\sum_\textbf{k}\left[\xi(\textbf{k})-E(\textbf{k})\right]+f\mu$. In Ref. [\onlinecite{CHZ}], the derivative $\partial f/\partial \mu$ is divergent for $\textbf{k}\rightarrow 0$ at $\mu=0$. The point $\textbf{k}=0$ can be eliminated in all summations over the momentum vectors by rewriting the free energy as  $F(\mu,\Delta)=\Delta^2/\lambda+1/N\sum_{\textbf{k}(\textbf{k} \neq 0)}\left[\xi(\textbf{k})-E(\textbf{k})\right]+\widetilde{f}\mu$, where the effective filling factor is $\widetilde{f}=f+\delta f$, and $\mu\delta f=\left[\xi(\textbf{k}=0)-E(\textbf{k}=0)\right]/N$. At a fixed filling factor $\widetilde{f}$,  the mean-field number- and gap-equations are: $\widetilde{f}=\frac{1}{N}\sum_{\textbf{k}(\textbf{k} \neq 0)}\left(1-
\frac{\xi(\textbf{k})}{E(\textbf{k})}\right)$, and   $1=\frac{\lambda}{N}\sum_{\textbf{k}(\textbf{k} \neq 0)}
\frac{\left(\cos k_x-\cos k_y\right)^2}{2E(\textbf{k})}$.
\section{Bethe-Salpeter equations for the collective modes of the extended Hubbard model}
The basic assumption in our BS formalism is that the
bound states of two fermions in the optical lattice at zero
temperature are described by the BS wave functions (BS
amplitudes). The BS amplitude determines the probability of finding the first fermion at  site $\textbf{r}_i$ with a spin $\sigma_1$ at
the moment $t_1$, and the second fermion at  site $\textbf{r}_j$ with a spin $\sigma_2$ at
the moment $t_2$. The BS amplitudes depend on the relative internal time $t_1 - t_2$ and on the "center-of-mass" time
$(t_1 + t_2)/2$: $\Phi^\textbf{Q}
_{\sigma_1,\sigma_2} (\textbf{r}_i, \textbf{r}_j ; t_1, t_2) =
e^{\imath [\textbf{Q.}(\textbf{r}_i + \textbf{r}_j)/2 - \omega(\textbf{Q})(t_1 + t_2)/2]}
\Psi^\textbf{Q}
_{\sigma_1,\sigma_2}(\textbf{r}_i - \textbf{r}_j , t_1 - t_2)$, where $\textbf{Q}$ and $\omega(\textbf{Q})$ are the collective-mode momentum and the corresponding dispersion, respectively. In the case of  the extended Hubbard model, the kernel $I$  of the BS equation $[K^{(0)-1}-I]\Psi=0$ is time independent; therefore, one can use the BS equation for the
equal-time BS amplitudes $\Psi^\textbf{Q}
_{\sigma_1,\sigma_2}(\textbf{r}_i - \textbf{r}_j ,0)$. The  free two-particle propagator $K^{(0)}$ in the BS equation
  is a product of two  fully dressed single-particle Green's functions, but in the GRPA the  fully dressed single-particle Green's functions were replaced by the corresponding Green's functions  $\left(%
\begin{array}{cc}
  G^{\uparrow,\uparrow}(\textbf{k},\textbf{Q},\omega)&G^{\uparrow,\downarrow}(\textbf{k},\textbf{Q},\omega)\\
  G^{\downarrow,\uparrow}(\textbf{k},\textbf{Q},\omega)&G^{\downarrow,\downarrow}(\textbf{k},\textbf{Q},\omega)
\end{array}%
\right)$ in the mean-field approximation. By introducing the functions $G^{\pm}(\textbf{k},\textbf{Q})$, the Fourier transforms of $\Psi^\textbf{Q}
_{\sigma_1,\sigma_2}(\textbf{r}_i - \textbf{r}_j ,0)$ can be written  as:
\begin{widetext}\begin{equation}\begin{split}& \Psi^{\downarrow,\uparrow}(\textbf{k},\textbf{Q})=\left[
(l_{\textbf{k},\textbf{Q}}+\gamma_{\textbf{k},\textbf{Q}})G^{+}(\textbf{k},\textbf{Q})+
(l_{\textbf{k},\textbf{Q}}-\gamma_{\textbf{k},\textbf{Q}})G^{-}(\textbf{k},\textbf{Q})\right]/2,\\&
\Psi^{\uparrow,\downarrow}(\textbf{k},\textbf{Q})=\left[
(l_{\textbf{k},\textbf{Q}}-\gamma_{\textbf{k},\textbf{Q}})G^{+}(\textbf{k},\textbf{Q})+
(l_{\textbf{k},\textbf{Q}}+\gamma_{\textbf{k},\textbf{Q}})G^{-}(\textbf{k},\textbf{Q})\right]/2,\\&
\Psi^{\uparrow,\uparrow}(\textbf{k},\textbf{Q})=\left[
(\widetilde{\gamma}_{\textbf{k},\textbf{Q}}-m_{\textbf{k},\textbf{Q}})G^{+}(\textbf{k},\textbf{Q})-
(m_{\textbf{k},\textbf{Q}}+\widetilde{\gamma}_{\textbf{k},\textbf{Q}})G^{-}(\textbf{k},\textbf{Q})\right]/2,\\&
\Psi^{\downarrow,\downarrow}(\textbf{k},\textbf{Q})=\left[
(\widetilde{\gamma}_{\textbf{k},\textbf{Q}}+m_{\textbf{k},\textbf{Q}})G^{+}(\textbf{k},\textbf{Q})+
(m_{\textbf{k},\textbf{Q}}-\widetilde{\gamma}_{\textbf{k},\textbf{Q}})G^{-}(\textbf{k},\textbf{Q})\right]/2.\nonumber
\end{split}\end{equation}
\end{widetext}
Here  the   form
factors are defined in the same manner  as in Ref. [\onlinecite{CG}]:
$\gamma_{\textbf{k},\textbf{Q}}=u_{\textbf{k}}u_{\textbf{k}+\textbf{Q}}+v_{\textbf{k}}v_{\textbf{k}+\textbf{Q}},\quad
l_{\textbf{k},\textbf{Q}}=u_{\textbf{k}}u_{\textbf{k}+\textbf{Q}}-v_{\textbf{k}}v_{\textbf{k}+\textbf{Q}},
\quad
\widetilde{\gamma}_{\textbf{k},\textbf{Q}}=u_{\textbf{k}}v_{\textbf{k}+\textbf{Q}}-u_{\textbf{k}+\textbf{Q}}v_{\textbf{k}},
$ and $ m_{\textbf{k},\textbf{Q}}=
u_{\textbf{k}}v_{\textbf{k}+\textbf{Q}}+u_{\textbf{k}+\textbf{Q}}v_{\textbf{k}}
$, where $u_\textbf{k}=\sqrt{\left[1+\xi_\textbf{k}/E(\textbf{k})\right]/2}$, and $v_\textbf{k}=\sqrt{\left[1-\xi_\textbf{k}/E(\textbf{k})\right]/2}$.

The functions $G^{\pm}(\textbf{k},\textbf{Q})$, as well as  the  collective-excitation energy $\omega(\textbf{Q})$,  are defined by the solutions of the following coupled equations:\cite{ZK}
\begin{widetext}\begin{equation}
\begin{split}
&[\omega(\textbf{Q})-\varepsilon(\textbf{k},\textbf{Q})]G^{+}(\textbf{k},\textbf{Q})=
-\frac{U}{2N}\sum_{\textbf{q}}
\left[\gamma_{\textbf{k},\textbf{Q}}\gamma_{\textbf{q},\textbf{Q}}+
l_{\textbf{k},\textbf{Q}}l_{\textbf{q},\textbf{Q}}\right]
G^{+}(\textbf{q},\textbf{Q})+\frac{U}{2N}\sum_{\textbf{q}}
\left[\gamma_{\textbf{k},\textbf{Q}}\gamma_{\textbf{q},\textbf{Q}}-
l_{\textbf{k},\textbf{Q}}l_{\textbf{q},\textbf{Q}}\right]G^{-}(\textbf{q},\textbf{Q})\\&-
\frac{1}{2N}\sum_{\textbf{q}}V(\textbf{k}-\textbf{q})
\left[\gamma_{\textbf{k},\textbf{Q}}\gamma_{\textbf{q},\textbf{Q}}+
l_{\textbf{k},\textbf{Q}}l_{\textbf{q},\textbf{Q}}+\widetilde{\gamma}_{\textbf{k},\textbf{Q}}
\widetilde{\gamma}_{\textbf{q},\textbf{Q}}+
m_{\textbf{k},\textbf{Q}}m_{\textbf{q},\textbf{Q}}\right]
G^{+}(\textbf{q},\textbf{Q})\\&+
\frac{1}{2N}\sum_{\textbf{q}}V(\textbf{k}-\textbf{q})
\left[\gamma_{\textbf{k},\textbf{Q}}\gamma_{\textbf{q},\textbf{Q}}-
l_{\textbf{k},\textbf{Q}}l_{\textbf{q},\textbf{Q}}+
\widetilde{\gamma}_{\textbf{k},\textbf{Q}}
\widetilde{\gamma}_{\textbf{q},\textbf{Q}}-m_{\textbf{k},\textbf{Q}}m_{\textbf{q},\textbf{Q}}\right]
G^{-}(\textbf{q},\textbf{Q})\\
&+\frac{U}{2N}\sum_{\textbf{q}}
\widetilde{\gamma}_{\textbf{k},\textbf{Q}}\widetilde{\gamma}_{\textbf{q},\textbf{Q}}
\left[G^{+}(\textbf{q},\textbf{Q})-G^{-}(\textbf{q},\textbf{Q})\right]
-\frac{U+2V(\textbf{Q})}{2N}\sum_{\textbf{q}}
m_{\textbf{k},\textbf{Q}}m_{\textbf{q},\textbf{Q}}
\left[G^{+}(\textbf{q},\textbf{Q})+G^{-}(\textbf{q},\textbf{Q})\right],
\label{BS1}
\end{split}
\end{equation}
\begin{equation}
\begin{split}
&[\omega(\textbf{Q})+\varepsilon(\textbf{k},\textbf{Q})]G^{-}(\textbf{k},\textbf{Q})=
\frac{U}{2N}\sum_{\textbf{q}}
\left[\gamma_{\textbf{k},\textbf{Q}}\gamma_{\textbf{q},\textbf{Q}}+
l_{\textbf{k},\textbf{Q}}l_{\textbf{q},\textbf{Q}}\right]
G^{-}(\textbf{q},\textbf{Q})-\frac{U}{2N}\sum_{\textbf{q}}
\left[\gamma_{\textbf{k},\textbf{Q}}\gamma_{\textbf{q},\textbf{Q}}-
l_{\textbf{k},\textbf{Q}}l_{\textbf{q},\textbf{Q}}\right]G^{+}(\textbf{q},\textbf{Q})\\&+
\frac{1}{2N}\sum_{\textbf{q}}V(\textbf{k}-\textbf{q})
\left[\gamma_{\textbf{k},\textbf{Q}}\gamma_{\textbf{q},\textbf{Q}}+
l_{\textbf{k},\textbf{Q}}l_{\textbf{q},\textbf{Q}}+\widetilde{\gamma}_{\textbf{k},\textbf{Q}}
\widetilde{\gamma}_{\textbf{q},\textbf{Q}}+
m_{\textbf{k},\textbf{Q}}m_{\textbf{q},\textbf{Q}}\right]
G^{-}(\textbf{q},\textbf{Q})\\&-
\frac{1}{2N}\sum_{\textbf{q}}V(\textbf{k}-\textbf{q})
\left[\gamma_{\textbf{k},\textbf{Q}}\gamma_{\textbf{q},\textbf{Q}}-
l_{\textbf{k},\textbf{Q}}l_{\textbf{q},\textbf{Q}}+
\widetilde{\gamma}_{\textbf{k},\textbf{Q}}
\widetilde{\gamma}_{\textbf{q},\textbf{Q}}-m_{\textbf{k},\textbf{Q}}m_{\textbf{q},\textbf{Q}}\right]
G^{+}(\textbf{q},\textbf{Q})\\
&+\frac{U}{2N}\sum_{\textbf{q}}
\widetilde{\gamma}_{\textbf{k},\textbf{Q}}\widetilde{\gamma}_{\textbf{q},\textbf{Q}}
\left[G^{+}(\textbf{q},\textbf{Q})-G^{-}(\textbf{q},\textbf{Q})\right]
+\frac{U+2V(\textbf{Q})}{2N}\sum_{\textbf{q}}
m_{\textbf{k},\textbf{Q}}m_{\textbf{q},\textbf{Q}}
\left[G^{+}(\textbf{q},\textbf{Q})+G^{-}(\textbf{q},\textbf{Q})\right],
\label{BS2}
\end{split}
\end{equation}
\end{widetext}
where $\varepsilon(\textbf{k},\textbf{Q})=E(\textbf{k}+\textbf{Q})+E(\textbf{k})$.

The first four terms in the right-hand sites (RHS) of the BS equations  represent the direct interactions in the BS kernel, while the last two terms in the RHS correspond to the exchange interactions. The existence of two U-dependent exchange terms is due to the fact that the Hubbard interaction  has a spin-dependent part and a spin-independent part.\cite{F} The spin-independent part of the interaction appears in combination with the spin-independent interaction $V(\textbf{Q})$ between  the atoms on the near-neighbor sites of the lattice.

As it is known,  the gauge invariance is restored by the existence of the Goldstone mode whose $\omega(\textbf{Q})$ energy approaches zero at $\textbf{Q}=0$. From the BS point of view, the  Goldstone theorem  corresponds to the so-called trivial solution of the BS equations:  $G^{+}(\textbf{k},\textbf{Q}=\textbf{0})=-G^{-}(\textbf{k},\textbf{Q}=\textbf{0})
=\Delta_\textbf{k}/2E(\textbf{k})=\Psi^{\downarrow,\uparrow}(\textbf{k},\textbf{Q}=\textbf{0})=
-\Psi^{\uparrow,\downarrow}(\textbf{k},\textbf{Q}=\textbf{0})$, $\Psi^{\uparrow,\uparrow}(\textbf{k},\textbf{Q}=\textbf{0})=\Psi^{\downarrow,\downarrow}(\textbf{k},\textbf{Q}
=\textbf{0})=0$. As it is  easy to see, the trivial solution  recovers the gap equation
(\ref{GGap}).

By setting the  near-neighbor interaction equal to zero, it is straightforward to prove  that the
existence of a nontrivial solution requires that the secular
determinant (\ref{BSU}) is equal to zero.

We now evaluate the collective-mode  energy by setting the  Hubbard interaction to zero. First, we set  $\Gamma_2(\textbf{k})=\Gamma_3(\textbf{k})=\Gamma_4(\textbf{k})=0$, and  neglect $\widetilde{\gamma}_{\textbf{k},\textbf{Q}}$ and $m_{\textbf{k},\textbf{Q}}$ terms. The
existence of a nontrivial solution requires that the following  secular determinant
 \begin{equation}
\left|
\begin{array}{cc}
\lambda^{-1}+ I_{\gamma,\gamma}^{1,1} &J_{\gamma,l}^{1,1}\\
J_{\gamma,l}^{1,1}&\lambda^{-1}+ I_{l,l}^{1,1}
\end{array}%
\right|=0\label{BS22}\end{equation} is equal to zero; therefore, the corresponding equation for the collective modes  is
\begin{equation}1+\lambda\left(I_{\gamma,\gamma}^{1,1}
+I_{l,l}^{1,1}\right)+\lambda^2\left[I_{\gamma,\gamma}^{1,1}I_{l,l}^{1,1}
-\left(J_{\gamma,l}^{1,1}\right)^2\right]=0,\label{GEq}\end{equation} where the quantities $I^{i,j}_{a,b}$ and $J^{i,j}_{a,b}$ are defined as follow:
\begin{equation}\begin{split}&
I^{i,j}_{a,b}=\frac{1}{N}\sum_\textbf{k}a_{\textbf{k},\textbf{Q}}b_{\textbf{k},\textbf{Q}}
\frac{\Gamma_i(\textbf{k})\Gamma_j(\textbf{k})\varepsilon(\textbf{k},\textbf{Q})}{\omega^2(\textbf{Q})-\varepsilon^2(\textbf{k},\textbf{Q})}
,\nonumber\\&
J^{i,j}_{a,b}=\frac{1}{N}\sum_\textbf{k}a_{\textbf{k},\textbf{Q}}b_{\textbf{k},\textbf{Q}}
\frac{\Gamma_i(\textbf{k})\Gamma_j(\textbf{k})\omega(\textbf{Q})}{\omega^2(\textbf{Q})-\varepsilon^2(\textbf{k},\textbf{Q})}
.\nonumber\end{split}\end{equation}
Here $\Gamma_0(\textbf{k})=1$, and $a_{\textbf{k},\textbf{Q}}$ and $b_{\textbf{k},\textbf{Q}}$ are one of the following form factors: $\gamma_{\textbf{k},\textbf{Q}},l_{\textbf{k},\textbf{Q}}, \widetilde{\gamma}_{\textbf{k},\textbf{Q}}$ and $m_{\textbf{k},\textbf{Q}}$.

 In Ref. [\onlinecite{CHZ}], the collective modes  across the BCS-BEC transition for d-wave pairing are obtained in the  Gaussian approximation using the following secular determinant:  $\left|
\begin{array}{cc}
M_{11}&M_{11}\\
M_{12}&M_{22}
\end{array}%
\right|$,   $M_{11}=1+(\lambda/2)\left(I_{\gamma,\gamma}^{1,1}+I_{l,l}^{1,1}-2J_{\gamma,l}^{1,1}\right)$, $M_{22}=1+(\lambda/2)\left(I_{\gamma,\gamma}^{1,1}+I_{l,l}^{1,1}+2J_{\gamma,l}^{1,1}\right)$,  $M_{12}=(\lambda/2)\left(I_{l,l}^{1,1}-I_{\gamma,\gamma}^{1,1}\right)$. This determinant provides the same equation as the BS equation (\ref{GEq}).  From the BS point of view, the  Gaussian approximation for the collective-mode dispersion and the BS formalism both provide the same dispersion, if the form factors $\widetilde{\gamma}_{\textbf{k},\textbf{Q}}$ and $m_{\textbf{k},\textbf{Q}}$ are equal to zero.
If all $\Gamma_i(\textbf{k})$ terms in the  nearest-neighbor interaction are taken into account, the collective-mode dispersion within the Gaussian approximation  is defined by the $8\times 8$ secular determinant $G_{8\times 8}$ given in the Appendix.

The existence of a nontrivial solution of the BS equations (\ref{BS1}) and (\ref{BS2}) beyond the Gaussian approximation ($\widetilde{\gamma}_{\textbf{k},\textbf{Q}}\neq 0$ and $m_{\textbf{k},\textbf{Q}}\neq 0$) leads to the following $17\times 17$ secular determinant
$$Z_{17\times 17}=\left|
\begin{array}{cc}
D_{16\times 16}&N^T_{1\times 16}\\
N_{1\times 16}&[2V(\textbf{Q})]^{-1}+I^{0,0}_{m.m}
\end{array}%
\right|.$$
The direct interactions in the BS kernel form  the $16\times 16$ block $D_{16\times 16}=\left|
\begin{array}{cc}
G_{8\times 8}&A_{8\times 8}\\
A_{8\times 8}&B_{8\times 8}
\end{array}%
\right|$. The rest of the elements of $Z_{17\times 17}$  are related to the existence of the exchange interaction in the BS kernel.
The blocks $A_{8\times 8}$ and $B_{8\times 8}$ and $N_{1\times 16}$ are defined  in the Appendix, and $T$ means transposed matrix operation.
\begin{figure}
\includegraphics[scale=0.7]{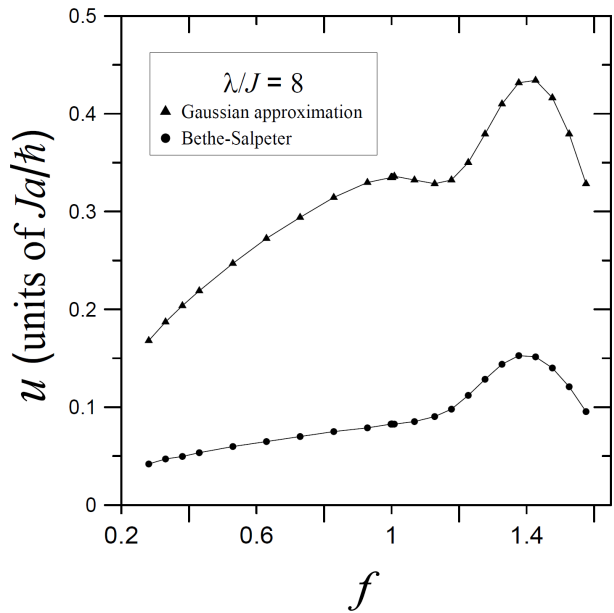}
  \caption{The speed of sound in 2D optical lattice
along the $(Q_x, 0)$ direction as a function of the filling factor $f$ for d-wave pairing calculated within the Gaussian approximation, and by solving the BS equations (solid
lines are guides to the eyes). The interaction strength is $\lambda=8J$, and the filling factor has been normalized such that $f=1$ at $\mu=0$. }\label{Fig4}\end{figure}
\begin{figure}\includegraphics[scale=0.7]{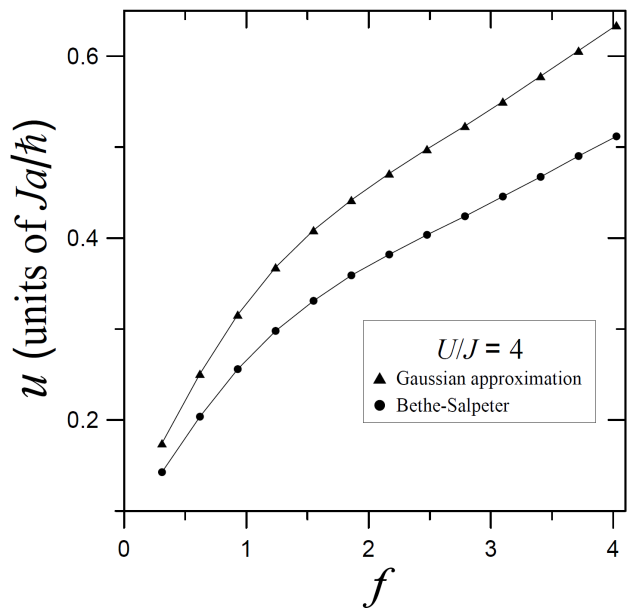}
  \caption{The speed of sound in 2D optical lattice
along the $(Q_x, 0)$ direction as a function of the filling factor $f$ for s-wave pairing calculated within the Gaussian approximation, and by  the BS formalism (solid
lines are guides to the eyes). The interaction strength is $U=4J$, and the filling factor has been normalized such that $f=1$ at $\mu=0$. }\label{Fig5}\end{figure}

In Fig.4, we have presented the numerical results of the speed of sound along the $(Q_x,0)$ direction as a function of the filling factor $f$.  The speed of sound in the Gaussian approximation is obtained using the equation $det|D_{8\times 8}|=0$, while  the $17\times 17$ secular determinant $Z_{17\times 17}$ is used in the BS formalism. As can be seen, the speed of sound across the BCS-BEC transition  is a smooth function.

\section{ Summary}
To summarize, we argued that the infrared divergence  predicted in Ref. [\onlinecite{CHZ}]  does not exist because in the BS formalism all elements of the secular determinant are defined only at points where $\textbf{Q}\neq 0$. At the point $\textbf{Q}=0$  the coupled BS equations are replaced by a single equation (the gap equation); therefore, the secular determinant does not exist, and there is no infrared divergence.

It is worth mentioning another important point, illustrated in Fig. 2 and Fig. 4. The comparison between the BS and the Gaussian methods leads to the conclusion that, in the case of s- and d-wave pairings,  the speed of sound calculated within the Gaussian approximation  is overestimated.  The use of the  Gaussian approximation  away from $f=1$ have to be justified by comparing it with the corresponding BS predictions. The results in Fig. 5  suggest that if the interaction decreases its strength, the difference between the Gaussian approach and the BS formalism becomes smaller.
\section{Acknowledgements}
This work was partially supported by the National Council for Science and Technology, CONACYT, Mexico, through the postdoctoral grant 259792.
\section{Appendix}\begin{widetext}In the case of d-wave pairing ($U=0$ and $\lambda>0$), the secular determinant in the Gaussian approximation is

\begin{equation}G_{8\times 8}=\left|
\begin{array}{cccccccc}
\lambda^{-1}+I^{1,1}_{\gamma,\gamma}&I^{1,2}_{\gamma,\gamma}&
I^{1,3}_{\gamma,\gamma}&I^{1,4}_{\gamma,\gamma}&J^{1,1}_{\gamma,l}&J^{1,2}_{\gamma,l}&
J^{1,3}_{\gamma,l}&J^{1,4}_{\gamma,l}\\
I^{1,2}_{\gamma,\gamma}&\lambda^{-1}+I^{2,2}_{\gamma,\gamma}&
I^{2,3}_{\gamma,\gamma}&I^{2,4}_{\gamma,\gamma}&J^{1,2}_{\gamma,l}&J^{2,2}_{\gamma,l}&
J^{2,3}_{\gamma,l}&J^{2,4}_{\gamma,l}\\
I^{1,3}_{\gamma,\gamma}&I^{2,3}_{\gamma,\gamma}&
\lambda^{-1}+I^{3,3}_{\gamma,\gamma}&I^{3,4}_{\gamma,\gamma}&J^{1,3}_{\gamma,l}&J^{2,3}_{\gamma,l}&
J^{3,3}_{\gamma,l}&J^{3,4}_{\gamma,l}\\
I^{1,4}_{\gamma,\gamma}&I^{2,4}_{\gamma,\gamma}&
I^{3,4}_{\gamma,\gamma}&\lambda^{-1}+I^{4,4}_{\gamma,\gamma}&J^{1,4}_{\gamma,l}&J^{2,4}_{\gamma,l}&
J^{3,4}_{\gamma,l}&J^{4,4}_{\gamma,l}\\
J^{1,1}_{\gamma,l}&J^{1,2}_{\gamma,l}&
J^{1,3}_{\gamma,l}&J^{1,4}_{\gamma,l}&\lambda^{-1}+I^{1,1}_{l,l}&I^{1,2}_{l,l}&
I^{1,3}_{l,l}&I^{1,4}_{l,l}\\
J^{1,2}_{\gamma,l}&J^{2,2}_{\gamma,l}&
J^{2,3}_{\gamma,l}&J^{2,4}_{\gamma,l}&I^{1,2}_{l,l}&\lambda^{-1}+I^{2,2}_{l,l}&
I^{2,3}_{l,l}&I^{2,4}_{l,l}\\
J^{1,3}_{\gamma,l}&J^{2,3}_{\gamma,l}&
J^{3,3}_{\gamma,l}&J^{3,4}_{\gamma,l}&I^{1,3}_{l,l}&I^{2,3}_{l,l}&
\lambda^{-1}+I^{3,3}_{l,l}&I^{3,4}_{l,l}\\
J^{1,4}_{\gamma,l}&J^{2,4}_{\gamma,l}&
J^{3,4}_{\gamma,l}&J^{4,4}_{\gamma,l}&I^{1,4}_{l,l}&I^{2,4}_{l,l}&
I^{3,4}_{l,l}&\lambda^{-1}+I^{4,4}_{l,l}
\end{array}%
\right|.\label{G88}\end{equation}
Beyond the Gaussian approximation, the following additional blocks appear in the secular determinant $Z_{17\times 17}$:
\begin{equation}A_{8\times 8}=\left|
\begin{array}{cccccccc}
I^{1,1}_{\gamma,\widetilde{\gamma}}&I^{1,2}_{\gamma,\widetilde{\gamma}}&
I^{1,3}_{\gamma,\widetilde{\gamma}}&I^{1,4}_{\gamma,\widetilde{\gamma}}&J^{1,1}_{l,\widetilde{\gamma}}&J^{1,2}_{l,\widetilde{\gamma}}&
J^{1,3}_{l,\widetilde{\gamma}}&J^{1,4}_{l,\widetilde{\gamma}}\\
I^{1,2}_{\gamma,\widetilde{\gamma}}&I^{2,2}_{\gamma,\widetilde{\gamma}}&
I^{2,3}_{\gamma,\widetilde{\gamma}}&I^{2,4}_{\gamma,\widetilde{\gamma}}&J^{1,2}_{l,\widetilde{\gamma}}&J^{2,2}_{l,\widetilde{\gamma}}&
J^{2,3}_{l,\widetilde{\gamma}}&J^{2,4}_{l,\widetilde{\gamma}}\\
I^{1,3}_{\gamma,\widetilde{\gamma}}&I^{2,3}_{\gamma,\widetilde{\gamma}}&
I^{3,3}_{\gamma,\widetilde{\gamma}}&I^{3,4}_{\gamma,\widetilde{\gamma}}&J^{1,3}_{l,\widetilde{\gamma}}&J^{2,3}_{l,\widetilde{\gamma}}&
J^{3,3}_{l,\widetilde{\gamma}}&J^{3,4}_{l,\widetilde{\gamma}}\\
I^{1,4}_{\gamma,\widetilde{\gamma}}&I^{2,4}_{\gamma,\widetilde{\gamma}}&
I^{3,4}_{\gamma,\widetilde{\gamma}}&I^{4,4}_{\gamma,\widetilde{\gamma}}&J^{1,4}_{l,\widetilde{\gamma}}&J^{2,4}_{l,\widetilde{\gamma}}&
J^{3,4}_{l,\widetilde{\gamma}}&J^{4,4}_{l,\widetilde{\gamma}}\\
J^{1,1}_{l,\widetilde{\gamma}}&J^{1,2}_{l,\widetilde{\gamma}}&
J^{1,3}_{l,\widetilde{\gamma}}&J^{1,4}_{l,\widetilde{\gamma}}&I^{1,1}_{l,m}&I^{1,2}_{l,m}&
I^{1,3}_{l,m}&I^{1,4}_{l,m}\\
J^{1,2}_{l,\widetilde{\gamma}}&J^{2,2}_{l,\widetilde{\gamma}}&
J^{2,3}_{l,\widetilde{\gamma}}&J^{2,4}_{l,\widetilde{\gamma}}&I^{1,2}_{l,m}&I^{2,2}_{l,m}&
I^{2,3}_{l,m}&I^{2,4}_{l,m}\\
J^{1,3}_{l,\widetilde{\gamma}}&J^{2,3}_{l,\widetilde{\gamma}}&
J^{3,3}_{l,\widetilde{\gamma}}&J^{3,4}_{l,\widetilde{\gamma}}&I^{1,3}_{l,m}&I^{2,3}_{l,m}&
I^{3,3}_{l,m}&I^{3,4}_{l,m}\\
J^{1,4}_{l,\widetilde{\gamma}}&J^{2,4}_{l,\widetilde{\gamma}}&
J^{3,4}_{l,\widetilde{\gamma}}&J^{4,4}_{l,\widetilde{\gamma}}&I^{1,4}_{l,m}&I^{2,4}_{l,m}&
I^{3,4}_{l,m}&I^{4,4}_{l,m}
\end{array}%
\right|,\label{A88}\end{equation}
\begin{equation}B_{8\times 8}=\left|
\begin{array}{cccccccc}
\lambda^{-1}+I^{1,1}_{\widetilde{\gamma},\widetilde{\gamma}}&I^{1,2}_{\widetilde{\gamma},\widetilde{\gamma}}&
I^{1,3}_{\widetilde{\gamma},\widetilde{\gamma}}&I^{1,4}_{\widetilde{\gamma},\widetilde{\gamma}}&J^{1,1}_{\widetilde{\gamma},m}&J^{1,2}_{\widetilde{\gamma},m}&
J^{1,3}_{\widetilde{\gamma},m}&J^{1,4}_{\widetilde{\gamma},m}\\
I^{1,2}_{\widetilde{\gamma},\widetilde{\gamma}}&\lambda^{-1}+I^{2,2}_{\widetilde{\gamma},\widetilde{\gamma}}&
I^{2,3}_{\widetilde{\gamma},\widetilde{\gamma}}&I^{2,4}_{\widetilde{\gamma},\widetilde{\gamma}}&J^{1,2}_{\widetilde{\gamma},m}&J^{2,2}_{\widetilde{\gamma},m}&
J^{2,3}_{\widetilde{\gamma},m}&J^{2,4}_{\widetilde{\gamma},m}\\
I^{1,3}_{\widetilde{\gamma},\widetilde{\gamma}}&I^{2,3}_{\widetilde{\gamma},\widetilde{\gamma}}&
\lambda^{-1}+I^{3,3}_{\widetilde{\gamma},\widetilde{\gamma}}&I^{3,4}_{\widetilde{\gamma},\widetilde{\gamma}}&J^{1,3}_{\widetilde{\gamma},m}&J^{2,3}_{\widetilde{\gamma},m}&
J^{3,3}_{\widetilde{\gamma},m}&J^{3,4}_{\widetilde{\gamma},m}\\
I^{1,4}_{\widetilde{\gamma},\widetilde{\gamma}}&I^{2,4}_{\widetilde{\gamma},\widetilde{\gamma}}&
I^{3,4}_{\widetilde{\gamma},\widetilde{\gamma}}&\lambda^{-1}+I^{4,4}_{\widetilde{\gamma},\widetilde{\gamma}}&J^{1,4}_{\widetilde{\gamma},m}&J^{2,4}_{\widetilde{\gamma},m}&
J^{3,4}_{\widetilde{\gamma},m}&J^{4,4}_{\widetilde{\gamma},m}\\
J^{1,1}_{\widetilde{\gamma},m}&J^{1,2}_{\widetilde{\gamma},m}&
J^{1,3}_{\widetilde{\gamma},m}&J^{1,4}_{\widetilde{\gamma},m}&\lambda^{-1}+I^{1,1}_{m,m}&I^{1,2}_{m,m}&
I^{1,3}_{m,m}&I^{1,4}_{m,m}\\
J^{1,2}_{\widetilde{\gamma},m}&J^{2,2}_{\widetilde{\gamma},m}&
J^{2,3}_{\widetilde{\gamma},m}&J^{2,4}_{\widetilde{\gamma},m}&I^{1,2}_{m,m}&\lambda^{-1}+I^{2,2}_{m,m}&
I^{2,3}_{m,m}&I^{2,4}_{m,m}\\
J^{1,3}_{\widetilde{\gamma},m}&J^{2,3}_{\widetilde{\gamma},m}&
J^{3,3}_{\widetilde{\gamma},m}&J^{3,4}_{\widetilde{\gamma},m}&I^{1,3}_{m,m}&I^{2,3}_{m,m}&
\lambda^{-1}+I^{3,3}_{m,m}&I^{3,4}_{m,m}\\
J^{1,4}_{\widetilde{\gamma},m}&J^{2,4}_{\widetilde{\gamma},m}&
J^{3,4}_{\widetilde{\gamma},m}&J^{4,4}_{\widetilde{\gamma},m}&I^{1,4}_{m,m}&I^{2,4}_{m,m}&
I^{3,4}_{m,m}&\lambda^{-1}+I^{4,4}_{m,m}
\end{array}%
\right|,\label{B88}\end{equation}
\begin{equation}
N_{1\times 16}=\left(J^{1,0}_{\gamma,m},J^{2,0}_{\gamma,m},J^{3,0}_{\gamma,m},J^{4,0}_{\gamma,m},
I^{1,0}_{l,m},I^{2,0}_{l,m},I^{3,0}_{l,m},I^{4,0}_{l,m},J^{1,0}_{\widetilde{\gamma},m},J^{2,0}_{\widetilde{\gamma},m},
J^{3,0}_{\widetilde{\gamma},m},J^{4,0}_{\widetilde{\gamma},m},I^{1,0}_{m,m},I^{2,0}_{m,m},I^{3,0}_{m,m},I^{4,0}_{m,m}\right).
\label{N116}\end{equation}

\end{widetext}


\begin{thebibliography}{999}
%
\bibitem{G1} J. R. Engelbrecht, M. Randeria, and C. A. R. S$\acute{a}$ de Melo, Phys. Rev. B
\textbf{55}, 15153 (1997).
%
\bibitem{Jap} Y. Yunomae, D. Yamamoto, I. Danshita, N.
Yokoshi, S. Tsuchiya, Phys. Rev. A 80
 063627 (2009).
%
\bibitem{BS} Z. Koinov, Ann. Phys. (Berlin) \textbf{522},  693  (2010).
%
\bibitem{Com} R. Combescot, M. Yu. Kagan, and S. Stringari, Phys. Rev.
A \textbf{74}, 042717 (2006).
%
\bibitem{CHZ} G. Cao, L. He, and P. Zhuang, Phys. Rev. A \textbf{87}, 013613 (2013).
%
\bibitem{UV} E. Dagotto, J. Riera, Y. C. Chen, A. Moreo, A. Nazarenko, F. Alcaraz and
F. Ortolani,  Phys. Rev. B \textbf{49}, 3548 (1994).

%
\bibitem{CG} R. Côté and A. Griffin, Phys. Rev. B \textbf{48}, 10404 (1993).
%
\bibitem{ZK} Z. G. Koinov, P. Nash, Phys. Rev. B \textbf{82}, 014528 (2010); Z. G. Koinov, Ann. Phys.
(Berlin) \textbf{524}, 421 (2012).
%
\bibitem{F} E. Fradkin, Field theories of condensed matter systems, Addison-Wesley (1991).


\end{thebibliography}
\end{document}